\documentclass{iaus}
\usepackage{graphicx}

\title[Galactic and Cosmic Chemical Evolution with Hypernovae] 
{Galactic and Cosmic Chemical Evolution \break with Hypernovae}

\author[Chiaki Kobayashi]   
{Chiaki Kobayashi}

\affiliation{Max-Planck-Institute for Astrophysics \break Karl-Schwarzschild-Str. 1, D-85741 Garching, Germany \break email: chiaki@MPA-Garching.MPG.DE}

\pubyear{2005}
\volume{228}  
\pagerange{1--7}
\date{?? and in revised form ??}
\jname{From Lithium to Uranium: Elemental Tracers of Early Cosmic Evolution}
\editors{V. Hill, P. Fran\c{c}ois \& F. Primas, eds.}
\begin{document}

\def\gtsim {>\kern-1.2em\lower1.1ex\hbox{$\sim$}~}   
\def\ltsim {<\kern-1.2em\lower1.1ex\hbox{$\sim$}~}   
\def \apj {ApJ}
\def \apjs {ApJS}
\def \aj  {AJ}
\def \aap {A\&A} 
\def \mnras {MNRAS}
\def \pasj {PASJ}
\def \araa {ARA\&A}

\maketitle

\begin{abstract}
We provide new nucleosynthesis yields depending on metallicity and energy (i.e., (normal supernovae and hypernovae), and show the evolution of heavy element abundances from C to Zn in the solar neighborhood.
We then show the chemodynamical simulation of the Milky Way Galaxy and discuss the G-dwarf problem.
We finally show the cosmological simulation and discuss the galaxy formation and chemical enrichment.
\end{abstract}

\firstsection 

\vspace*{-1mm}

\section{Introduction}

Elemental abundance ratios are the treasure house of information on star formation and galaxy formation because different types of supernovae produce different heavy elements with different timescales.
High resolution spectroscopy gives the elemental abundance patterns of individual stars in our Galaxy and nearby dwarf spheroidal galaxies, and of high-redshift quasar absorption line systems.
These observations have shown different abundance patterns, which suggest different chemical enrichment histories of these objects and probably different initial mass functions (IMF) depending on environment.
However, to discuss these issues, nucleosynthesis yields have involved too many inconsistencies with recent observations.
Umeda \& Nomoto (2002 and 2005, hereafter UN05 and UN02) updated the progenitor star models with a metallicity range of $Z=0-0.02$, and calculated the explosive nucleosynthesis yields with larger energy than $10^{51}$ erg (1 foe).
They showed some of their yields are in good agreement with the observed abundance patterns of some EMP stars (UN05).
In this paper, we calculate a new yield table for ``typical'' SNe II and HNe, including the dependences of the explosion energy and metallicity (\S 2).
We show the evolution of elemental abundance ratios from carbon to zinc in the solar neighborhood using a one-zone chemical evolution model (\S 3).
We then apply our yields to a chemodynamical model of the Milky Way Galaxy with GRAPE (\S 4) and the cosmological simulation with GADGET (\S 5).

\vspace*{-3mm}

\section{Nucleosynthesis Yields}

Our calculation method is the same as UN05, and the details are described in Umeda et al. (2000).
From the light curve and spectra fitting of individual supernovae, the mass of progenitors stars, explosion energy, and produced $^{56}$Ni mass (witch decayed to $^{56}$Fe) have been obtained (e.g., Nomoto et al. 2002 for a review).
There exist two distinct types of core-collapse supernovae.
We set two mass-energy relations; a constant $E_{51}=1$ for normal SNe II, and $E_{51}=10,10,20$, and $30$ respectively for $20$, $25$, $30$, and $40M_\odot$ HNe.

For a given progenitor model, if the explosion mechanism is specified (or the procedure for the artificial explosion like Woosley \& Weaver (1995) and Limongi \& Chieffi (2003)), the remnant mass is uniquely determined as a function of the explosion energy. However, we do not specify the explosion mechanism and treat the mixing and fallback as free-parameters, especially because precise explosion mechanism is unknown for hypernovae.
The mass-cut is determined to give $M({\rm Fe})\sim0.1M_\odot$ for SNe II.
For HNe, the mixing and fallback parameters are determined to give [O/Fe] $=0.5$.

Two tables of the nucleosynthesis yields for SNe II and HNe will be shown in Kobayashi et al. (2005, in preparation).
The abundance ratios relative to solar iron abundance, [X/Fe], are shown in Nomoto et al. (2005, in this proceedings).
The yield masses of $\alpha$ elements (O, Ne, Mg, Si, S, Ar, Ca, and Ti) are larger for more massive stars because of the larger mantle mass. 
For SNe, the abundance ratio [$\alpha$/Fe] is larger for more massive stars.
The abundances of odd-Z elements (Na, Al, P, ...) strongly depend on the metallicity.
[Na/Fe] and [Al/Fe] of metal-free stars are smaller by $\sim 1.0$ and $0.7$ dex than solar abundance stars.

\vspace*{-1mm}

\section{Chemical Evolution of the Solar Neighborhood}

\begin{figure}
\centerline{
\scalebox{0.5}{
\includegraphics{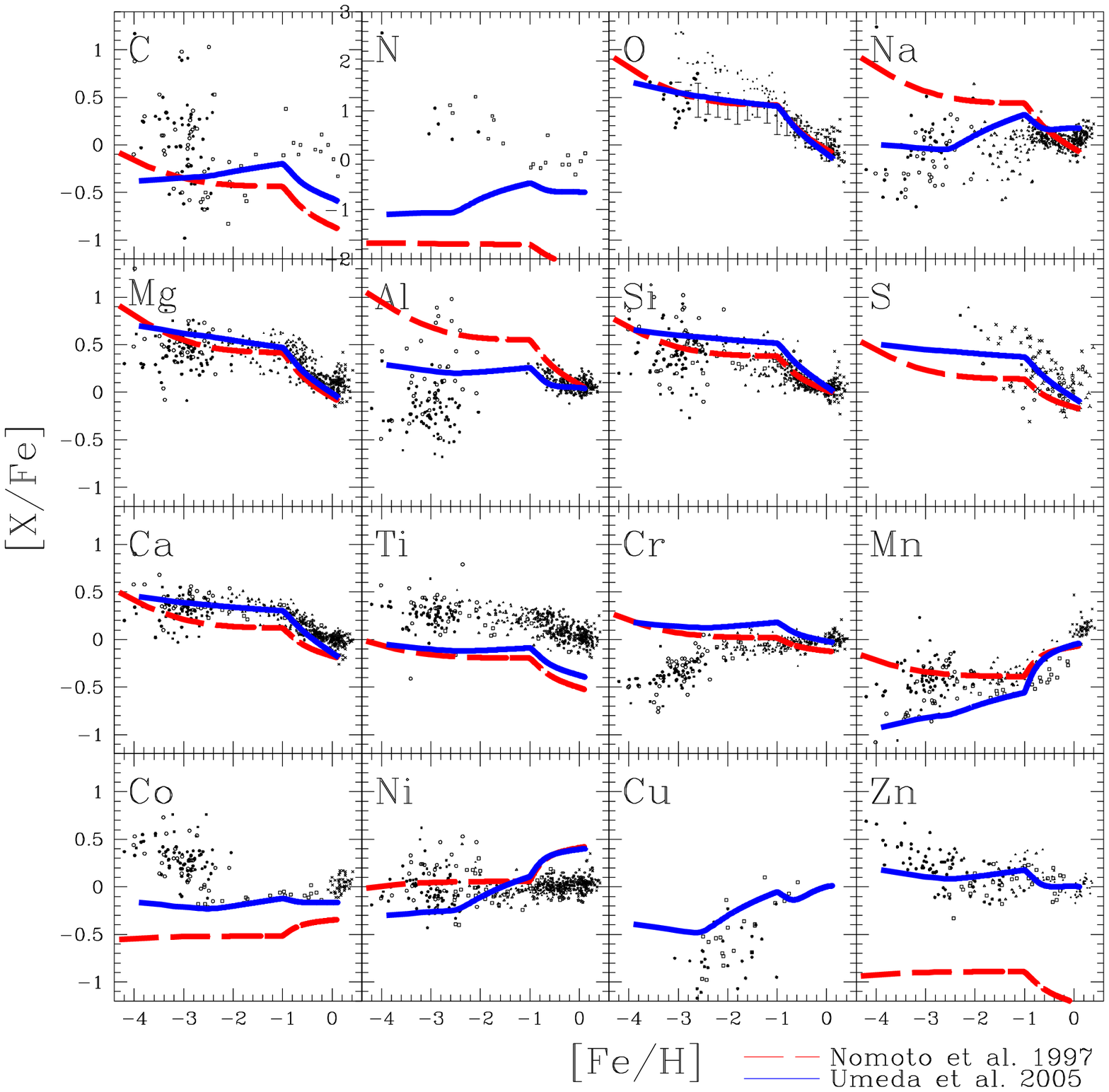}
}
}
\caption{\label{fig:xfe}
[X/Fe]-[Fe/H] relations for the model with our yields (solid line) and K98 model with N97 yield (dashed line).
Observational data sources are \cite{edv93}, \cite{mcw95}, \cite{rya96}, \cite{mel02}, \cite{gra03}, \cite{ben03}, \cite{cay04}, \cite{hon04}, and so on.
}
\end{figure}

In the chemical evolution model, we should introduce one important parameter to describe the fraction of hypernovae, $\epsilon_{\rm HN}$. 
Although the mass-energy relation has been obtained from the light curve fitting of individual supernovae, at the moment there is no constraint on the energy distribution function because of the poor statistics.
$\epsilon_{\rm HN}$ may depend on metallicity, and may be constrained with the gamma-ray burst rate.
Here we adopt $\epsilon_{\rm HN}=0.5$ independent of the mass and metallicity, and this gives good agreement of [$\alpha$/Fe] plateau against [Fe/H] (Fig. \ref{fig:xfe}).
We should note that the plateau value depend on the IMF, specifically on the slope $x$ and the upper limit $M_{\rm u}$.

We use the chemical evolution model that allows the infall of material from
outside the disk region (see Kobayashi et al. 2000, hereafter K00, for the formulation).
We adopt the Galactic age 13 Gyr, infall timescale $5$ Gyr, star formation coefficient $0.45$ Gyr$^{-1}$, and the Salpeter IMF with a slope of $x=1.35$ in the range of $0.07M_\odot \leq M \leq 50M_\odot$.
The treatment of SNe Ia is the same as in K00 but with [$b_{\rm RG}=0.02, b_{\rm MS}=0.04$].
The metallicity inhibition of SNe Ia at [Fe/H] $\leq -1.1$ is included (Kobayashi et al. 1998, hereafter K98).
These parameter can be determined uniquely from the metallicity distribution function (MDF) and the [O/Fe]-[Fe/H] evolutionary trend at [Fe/H] $\gtsim-1$ in the solar neighborhood.

Figures \ref{fig:xfe} shows the evolutions of heavy element abundance ratios [X/Fe] against [Fe/H]. The solid and dashed lines are for the model with our yields and K98 model with Nomoto et al. (1997, hereafter N97)'s yields, respectively.

{\it Oxygen ----}
Although O is the most abundant heavy element that covers the half of metallicity for the solar abundance and is one of the best described elements in nucleosynthesis, the observation bas been debated. 
However, when a suitable temperature scale is adopted and the NLTE and 3D effects are taken into account, [OI], OI, and IR OH lines give consistent results.
In our model, [O/Fe] is $0.41$ at [Fe/H] $=-1$ and slightly increases to $0.55$ at [Fe/H] $=-3$, being consistent with these observations (except for UV OH results, e.g., Israelian et al. 1998).
The gradual increase of [O/Fe] with decreasing [Fe/H] is caused by massive, metal-poor SNe II, and HNe.
The metallicity dependence is as small as $0.2$ dex between $Z=Z_\odot$ and $Z=0$. The mass dependence is larger, but such dependence is weaken because HNe produce more Fe.
From [Fe/H] $\sim -1$, [O/Fe] decreases quickly due to much Fe production by SNe Ia (see K98 for SN Ia models).

{\it Magnesium ----}
Mg is one of the best observed elements with several lines and little NLTE effect.
Cayrel et al. (2004, hereafter C04) claimed that [Mg,Si,Ca,Ti/Fe] is constant as $\sim 0.2-0.3$ with a very small dispersion of $\sim 0.1$ dex.
In our model, [Mg/Fe] is $0.48$ at [Fe/H] $\sim-1$ and slightly increases to $0.62$ at [Fe/H] $\sim-3$, which is larger than $0.27$ in C04, but in good agreement with the observed relation over the wide range of [Fe/H].
For HNe, since we determine our parameters of mixing-fallback to give a constant [O/Fe], [Mg/Fe] is also close to $0.5$.
SNe II with $E_{51}=1$ typically provide [Mg/Fe] $\sim0.5$ with a variety from $-0.2$ ($Z=Z_\odot$) to $1$ ($40M_\odot$).
Therefore, the scatter of [Mg/Fe] can be small independent of the mixing process of interstellar medium.

{\it Silicon and Sulfur ----}
Observed Si abundance is represented by only two lines and affected by the contamination.
[Si/Fe] is $0.52-0.60$ for $-3\ltsim$ [Fe/H] $\ltsim-1$ in our model, which is a bit larger than $0.37$ in C04 and the other observations.
For S, because of the hardness of observation, the plateau value is unknown, and our prediction is [S/Fe] $=0.37-0.45$.
Some observations suggest a sharp increase decreasing [Fe/H] (Israelian \& Reboro 2001).
Theoretically, however, it is difficult to change the Si/S evolution.

{\it Calcium and Titanium ----}
Our model succeed in reproducing the observation with a plateau [Ca/Fe] $\sim0.31-0.39$, which is increased by $\sim 0.2$ dex than N97.
However, Ti is $\sim 0.4$ underabundant overall, which cannot be improved by changing our parameters.
Possible solution to increase Ti is a jet-like explosion with high temperature (Maeda \& Nomoto 2003).

{\it Sodium, Aluminum, and Copper ----}
The NLTE effect for Na and Al is large for metal-poor stars, and the observational data at [Fe/H] $\ltsim -2$ are shifted by a constant of $-0.2$ and $+0.5$, respectively (Frebel et al. 2005).
For Na, Al, and Cu, our models are almost consistent with the observations, although $\sim 0.3$ dex offsets remain.
The decreasing trend toward lower [Fe/H] is seen weakly because the mass dependence also affect in our one-zone model.
At [Fe/H] $\gtsim -1$, the U-shape track is well reproduced with the combination of SNe Ia and metal-dependent SNe II.

{\it Chromium, Manganese, Cobalt, and Nickel ----}
McWilliam et al. (1995) found the decreasing trend of [(Cr,Mn)/Fe] and the increasing trend of [Co/Fe] toward lower metallicity.
UN05 explained these by the energy effect where larger energy increases Co and decreases Cr and Mn, i) if the interstellar medium is not mixed and the EMP stars enriched only by single supernovae (Audouse \& Silk 1995) and ii) if the hydrogen mass swept by supernovae ejecta is proportional to the explosion energy.
However, C04 claimed that no relation is found in [Mn/Fe].
Here we are showing the average yields for typical SNe II and HNe, and such trends are not realized.
We discuss whether our mean values meet the observation at [Fe/H] $\sim-2$.
As a whole, our yields are better agreement with observations than N97 yields, although $\sim 0.1$ dex offsets still remain.
From [Fe/H] $\sim-1$ SNe Ia contributes, which is significantly appears as the [Mn/Fe] increase toward higher metallicity. 
This is confirmed both observationally and theoretically, and Mn can be a key element to discuss SNe Ia contribution, HN fraction, IMF, and so on.
The Ni overproduction from SNe Ia can be reduced (Iwamoto et al. 1999).

{\it Zinc ----}
The advantage of our yields is in Zn, which is an important element observed in damped Ly$\alpha$ (DLA) systems without the dust depletion effect.
Because of the large energy and our mixing-fallback treatment, Zn that is formed in the incomplete Si-burning region at the center of the progenitor can be enough produced and ejected.
At $-2\ltsim$ [Fe/H] $\ltsim-1$, [Zn/Fe] is constant to be $\sim 0.1$, and mildly decreases to $\sim0$ from [Fe/H] $\sim-1$ due to SNe Ia, which are consistent with observations.
Primas et al. (2000) found the increasing [Zn/Fe] trend toward lower metallicity.
UN05 explained this by the energy effect as well as iron-peak elements.
Such trend is not realized in our model.

We should note that the under-abundance of C and N is because we do not include the yields from AGB stars and stellar winds, which are dominant source of these elements.
K, Sc, Ti, and V are under-abundant, which cannot be increased changing our parameters. 
A low-density model (UN05) may increase Sc abundance.

\vspace*{-1mm}

\section{Chemodynamical Evolution of the Milky Way Galaxy}

In reality, the inter-stellar medium is not well mixed especially at the early stage of the galaxy formation.
We simulate the chemodynamical evolution of the Milky Way Galaxy, using the special purpose computer GRAPE for gravity and the SPH method for hydrodynamics
(see Kobayashi, Nakasato, \& Nomoto 2005, in preparation).
We introduce the various physical processes associated with the formation of stellar systems to construct the self-consistent three-dimensional chemodynamical models; radiative cooling, star formation, feedback of SNe II and Ia, and stellar winds, chemical enrichment, and the UV background radiation.
The details of our code are described in Kobayashi (2004, 2005).
The star formation parameter $c=0.1$ and the Salpeter IMF $x=1.35$ are adopted.

As the initial condition, we use the CDM initial fluctuation, which is generated by the COSMICS package by E. Bertschinger. We simulate the 3-$\sigma$ over dense region with the radius of $65$ kpc, the mass of $\sim 10^{12} M_\odot$ (baryon fraction of $0.1$), and $\sim 60000$ particles (the half for gas and the rest for dark matter). The initial angular momentum is given as rigid rotation with the spin parameter of $\lambda \sim 0.1$.
A galaxy form through the successive merging of subgalaxies with various masses. We choose an initial condition where the galaxy does not undergo major mergers, otherwise no spiral galaxy form.

The merging of sub-galaxies induces the initial star burst and the bulge forms by $z \gtsim 3$.
The gas accrete on the disk and disk stars formed at $1 \ltsim z \ltsim 3$.
The bulge has the de Vaucouleurs surface brightness profile with the effective radius of $1.5$ kpc, and the disk has the exponential surface brightness with $5$ kpc. 
In Figure 2a, we reproduce the age-metallicity relation. [Fe/H] increases to $\sim 0$ at $t \sim 2$ Gyr, because the merging of sub-galaxies induces the strong initial star burst.
In Figure 2b, we reproduce the [O/Fe]-[Fe/H] relation. [O/Fe] decreases because of the delayed iron enrichment of SNe Ia. If we do not include the metallicity effect on SNe Ia, we cannot reproduce the plateau at [Fe/H]$\ltsim -1$. 
The scatter is large at [Fe/H]$\gtsim -1$. This may be because the mixing of heavy elements among gas particles has not been included in our chemodynamical model.

Figure 2c shows the metallicity distribution functions. For the bulge (dashed line), the simulated distribution is in good agreement with the observation. For the solar neighborhood (solid line), the peak is consistent but the number of metal poor stars is still a bit larger than the observation (i.e., the G-dwarf problem). Since we include the effect of the UV background radiation, the initial star burst is smaller than the case without it, and the number of metal-poor stars is reduced. For the halo (dotted line), the peak is higher than the observation.
This is because the star burst occurs at the very early epoch before the gas accrete onto the disk. 
There may be room to improve the modeling such as the star formation criteria and the metal-dependent IMF.

\begin{figure}
\centerline{
\scalebox{0.45}{
\includegraphics[angle=-90]{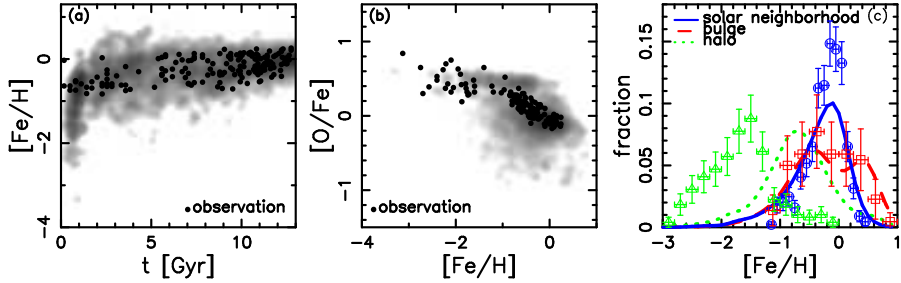}
}
}
\caption{(a) Age-metallicity relation and (b) [O/Fe]-[Fe/H] relation in the solar neighborhood. The contours show the mass density for the simulation. Observational data (dots) sources are in K98.
(c) Metallicity distribution functions in the bulge (dashed line), halo (dotted line), and solar neighborhood (solid line). Observational data for the solar neighborhood (circles), bulge (squares), and halo (triangles) are taken from Edvardsson et al. (1993), McWilliam \& Rich (1994), and Chiba \& Yoshii (1998), respectively.}
\end{figure}


\section{Cosmic Chemical Enrichment}

We simulate the evolution of a cosmological field including star formation, supernova feedback, and chemical enrichment with our new yields
(see Kobayashi, Springel, \& White 2005, in preparation).
We use a SPH code GADGET2 by Springel et al. (2000), and introduce the metal-dependent cooling (Sutherland \& Dopita 1993) and the chemical enrichment model by Kobayashi (2004).
The existence of hypernovae affect not only chemical enrichment but also feedback to suppress star formation.
Since the existence of heavy elements enhance gas cooling, this works complexly.
A self-consistent chemodynamical model is required.

We focus on a $\lambda$CDM cosmological model with parameters $H_0=70$ km s$^{-1}$ Mpc$^{-1}$, $\Omega_m=0.3$, $\Omega_\Lambda=0.7$, $\Omega_{\rm b}=0.04$, $n=1$, and $\sigma_8=0.9$.
The initial condition is calculated with GenIC in a $10 h^{-1}$ Mpc cubic box with periodic boundary conditions. 
Here we show the results with $N_{\rm DM}=N_{\rm gas}=54^3$.
The mass of a dark matter and gas particle is $5 \times 10^8 M_\odot$ and $7 \times 10^7 M_\odot$, respectively.

\begin{figure}
\centerline{
\scalebox{0.19}{
\includegraphics{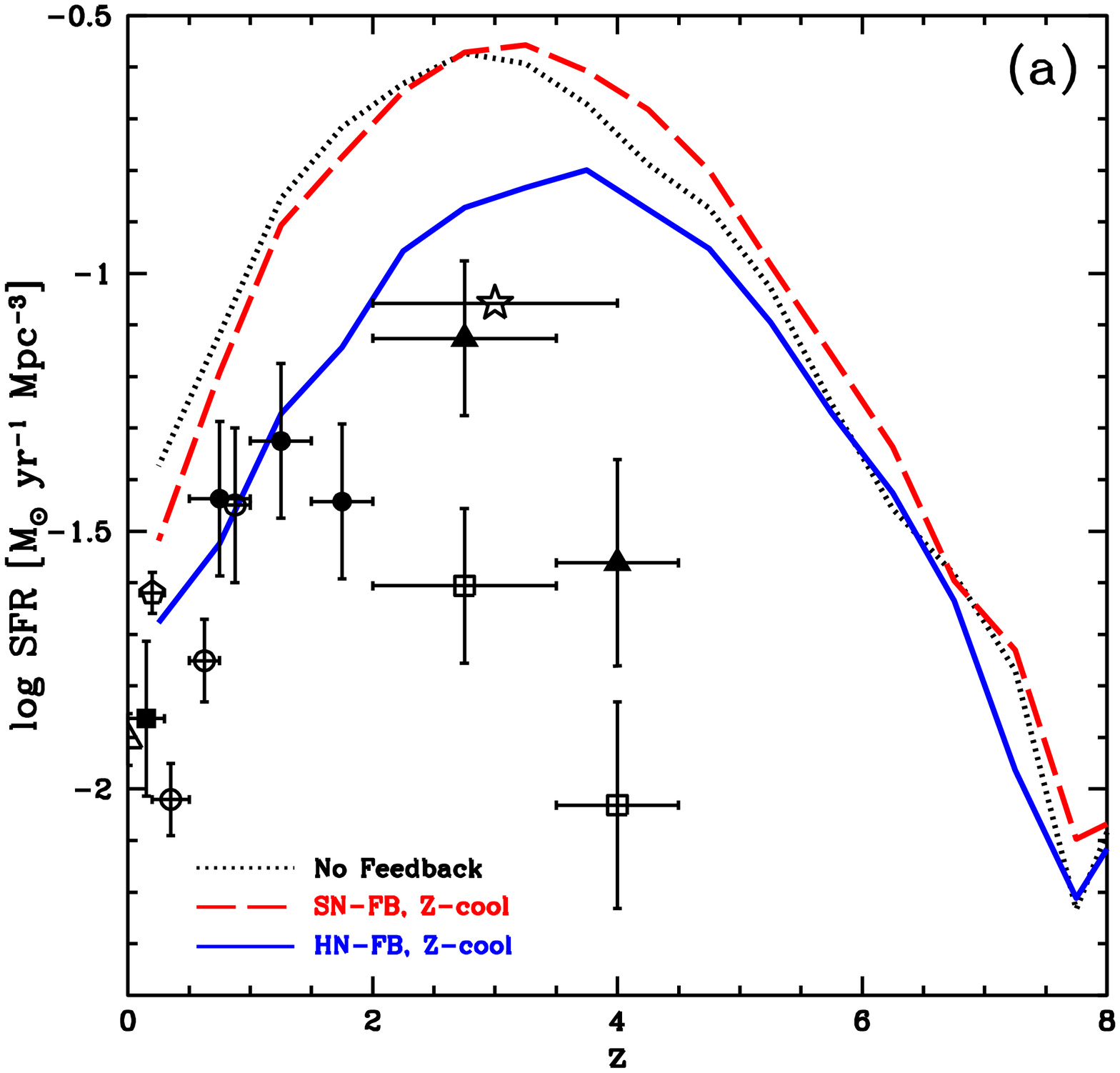}
\includegraphics{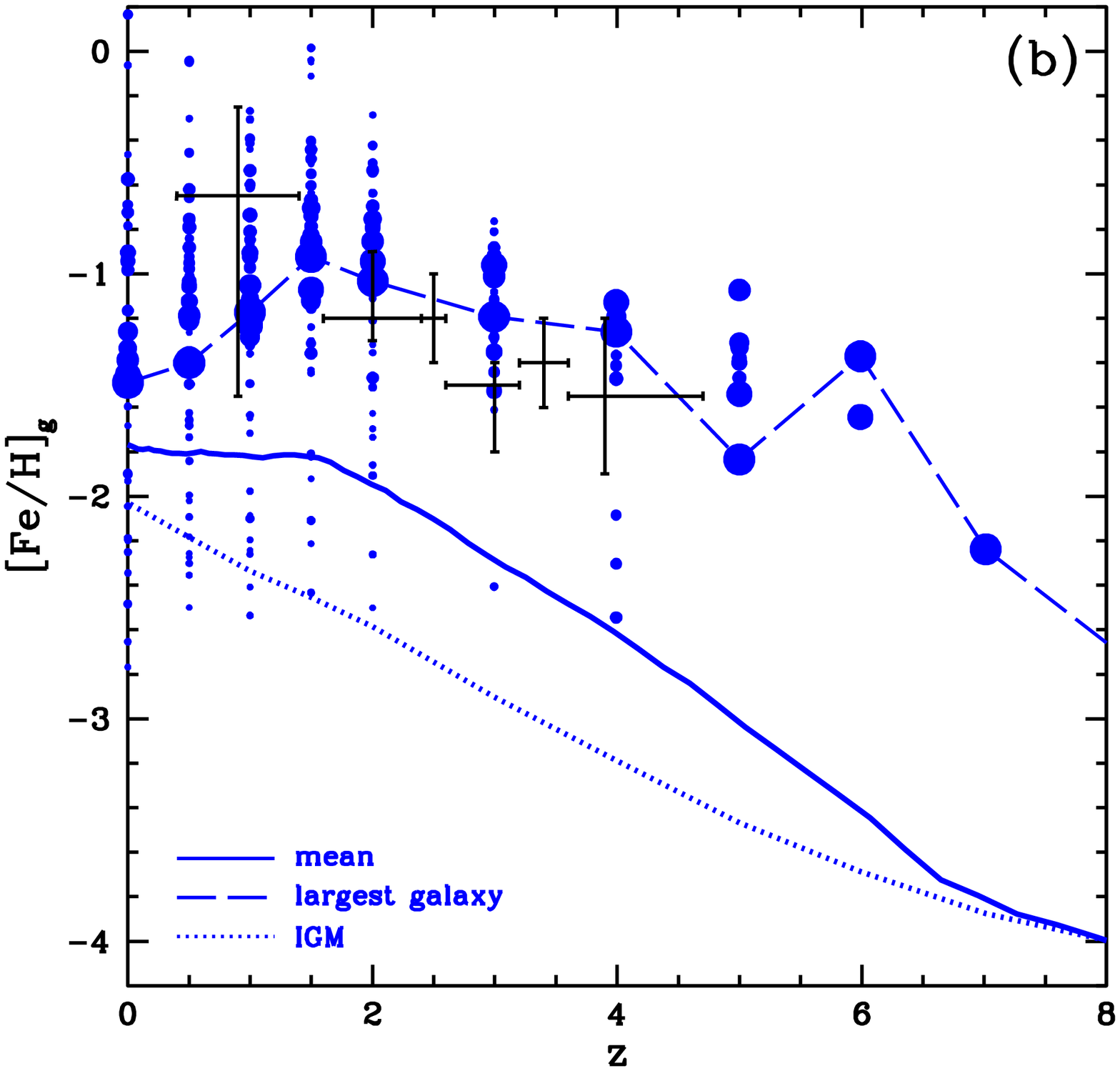}
\includegraphics{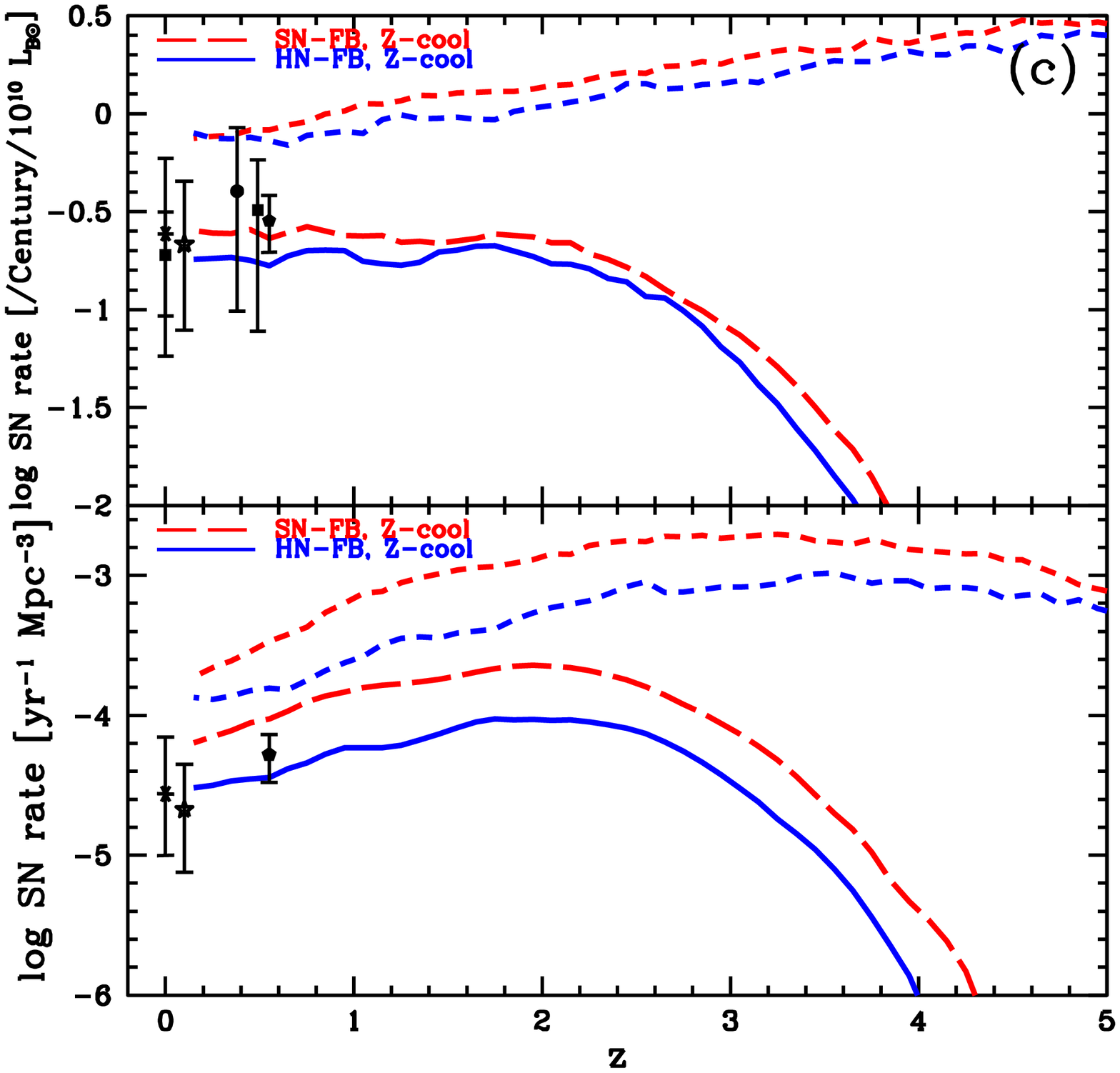}
}
}
\caption{\label{fig:sfr}
(a)
Cosmic star formation rates for no feedback (dotted line), the SN (dashed line) and HN feedback (solid line) cases with the metal-dependent cooling,.
Observational data (dots) sources are in K00.
(b)
Redshift evolution of iron abundances of gas.
The points show individual galaxies with the size representing the size of galaxies.
The solid, dashed, and dotted lines are for the mean, the largest galaxy in the simulated region, and the IGM, respectively.
The errorbars show the observations taken for the DLA systems (Prochaska et al. 2003),
(c)
The cosmic supernova rates in the SN unit (upper panel) and in the volume (lower panel) as a function of redshift.
The solid and dashed lines are for SNe Ia and II (including Ib, Ic, and HN), respectively.
The observational data sources are:
triangles, Cappellaro et al. (1999); star, Hamuy \& Pinto (1999); asterisk, Hardin et al. (2000);
circle, Pain et al. (1996); square, Reiss (2000); and pentagon, Pain et al. (2002).
}
\end{figure}

Figure 3a shows the cosmic star formation rates (SFR). 
The supernova feedback decreases the SFR from $z\sim3$, and the SFR is smaller by a factor of $2$ at $0\ltsim z \ltsim2$ than no feedback case (dotted line). 
However, the metal-dependent cooling increases the SFR as large as no feedback case (dashed line).
If we include the HN feedback (solid line), the SFR starts to be suppressed from $z\sim6$, and is smaller by a factor of $3$ at $0\ltsim z \ltsim3$. 
This is in good agreement with the observation.
With HNe, more metals are ejected with larger energy, and enriched hot gas can remain not forming stars, even with the metal-dependent cooling.

When and where stars form? In dwarf galaxies before they merge to massive galaxies. Thus stellar population in massive galaxies is as old as 10 Gyr.
From less massive galaxies, the galactic winds blow, which eject heavy elements into the inter-galactic medium (IGM) in a short timescale.
Figure 3b shows the evolution of iron abundance in gas phase.
Metal enrichment timescale depends on the environment.
On the average (solid line), [Fe/H] reaches $-3$ at $z \sim 5$ but at $z \sim 3$ in the IGM (dotted line).
The points show the abundances in individual galaxies identified by the FOF algorithm, which are consistent with the observations in the DLA systems (Prochaska et al. 2003).
In large galaxies, enrichment takes place so quickly that [Fe/H] reaches $\sim -2$ at $z\sim6$.
Smaller galaxies evolve with various timescales, and a large scatter is seen in $-2.5 \ltsim$[Fe/H]$\ltsim 0$ at present.

Because of the gas accretion, [$\alpha$/Fe] is as low as 0 even at [Fe/H]$\ltsim -1$, although we include the SN Ia metallicity effect.
In galaxies, the metallicity of the cold gas shows a relation between the galaxy mass, where the slope is consistent with the observation (Tremonti et al. 2004) but the zero-point offset is seen.
For the stellar population, the observed mass-metallicity relation (Kobayashi \& Arimoto 1999) can be reproduced, and these relation will be found since $z \sim 5$.
At present, however, star formation has not terminated in massive galaxies, and not many dwarf galaxies are forming.
The stellar population in massive galaxies have low [$\alpha$/Fe], and no relation is found between the galaxy mass.
This may be due to the lack of resolution, or may suggest that non-supernova feedback such as AGN is important.

Figure 3c show the cosmic supernova rate history.
With the HN feedback, the lower SFR results in lower SN II and Ia rates. 
The slower metal enrichment delay the SN Ia starting redshift.
The simulated SN Ia rate is in good agreement with the observations.
At $z \ltsim 2$, the cosmic SN Ia rate is constant in the unit of luminosity (SNu), and slowly increase in the unit of volume.
Because both of the lifetime and metallicity effects, the SN Ia rate gradually decreases from $z\sim2.5$, and drop off at $z\sim4$.
These observations can give some constraints on the HN efficiency, $\epsilon_{\rm HN}$.

\end{document}